\documentclass[12pt,preprint]{aastex}
\usepackage{rotating}
\usepackage{epsfig}

\begin{document}

\title{Star formation in the massive ``starless'' infrared dark cloud G0.253$+$0.016}

\author{Luis F. Rodr\'\i guez\altaffilmark{1,2} and Luis A. Zapata\altaffilmark{1}}

\altaffiltext{1}{Centro de Radioastronom\'\i a y Astrof\'\i sica, 
UNAM, Apdo. Postal 3-72 (Xangari), 58089 Morelia, Michoac\'an, M\'exico}
\altaffiltext{2}{Astronomy Department, Faculty of Science, King Abdulaziz University, 
P.O. Box 80203, Jeddah 21589, Saudi Arabia}

\email{lrodriguez,lzapata@crya.unam.mx}
 
\begin{abstract}
G0.253$+$0.016 is a remarkable massive infrared dark cloud located
within $\sim$100 pc of the galactic center. With a high mass of
$1.3 \times 10^5~M_\odot$, a compact average radius of $\sim$2.8 pc
and a low dust temperature of 23 K, it has been believed to be
a yet starless precursor to a massive Arches-like stellar cluster. 
We present sensitive JVLA 1.3 and 5.6 cm radio continuum observations 
that reveal the presence on three compact thermal radio sources projected
against this cloud. These radio sources are interpreted as HII regions
powered by $\sim$B0.5 ZAMS stars. We conclude that although
G0.253$+$0.016 does not show evidence of O-type star formation, there
are certainly early B-type stars embedded in it.
We detect three more sources in the periphery of G0.253$+$0.016
with non-thermal spectral indices. We suggest that these sources
may be related to the galactic center region and deserve further study.

\end{abstract}  

\keywords{
stars: pre-main sequence  --
ISM: jets and outflows -- 
ISM: individual: (G0.253$+$0.016) --
stars: radio continuum 
}

\section{Introduction}
The handful of young massive ($M \simeq 10^4-10^5~M_\odot$) stellar clusters
known in our Galaxy (Portegies Zwart et al. 2010) 
are thought to be a link between the more common open clusters
($M < 10^4~M_\odot$)
and the globular clusters and the extreme extragalactic super star
clusters ($M > 10^5~M_\odot$). Studying star formation in massive
cold molecular clouds that can be considered massive proto-clusters,
that is precursors to
young massive stellar clusters, is expected to shed light on
the process at these ``intermediate'' mass ranges.  
Recently, Ginsburg et al. (2012) investigated 
18 massive proto-clusters in the first Galactic quadrant outside of the
central kpc and found that all presented active star formation.

An apparent exception to this situation is the massive infrared dark cloud
G0.253$+$0.016, that was not considered in the Ginsburg et al.
(2012) study since it is located within 100 pc of the Galactic center.
This cloud has been observed in detail (see Longmore et al. 2012
for a review) and is considered to be starless or nearly starless,
with only a weak water maser detected at position 
$\alpha(2000) = 17^h~ 46^m~ 10\rlap.^{s}60 ; 
\delta(2000) = -28^\circ~ 42'~ 17\rlap.{''}5$, 
most likely associated with a deeply embedded 
intermediate or low-mass young stellar object
(Lis et al. 1994; Immer et al. 2012).

In this paper we report sensitive JVLA observations of G0.253$+$0.016
that reveal the presence of three radio continuum sources
apparently embedded in it. We also discuss the nature of four other radio sources detected in
the surroundings of G0.253$+$0.016.

\section{Observations}
The observations were made with the Karl G. Jansky Very Large Array 
of NRAO\footnote{The National 
Radio Astronomy Observatory is a facility of the National Science Foundation operated
under cooperative agreement by Associated Universities, Inc.} centered at rest 
frequencies of 20.943 (1.3 cm) and 5.307 (5.6 cm) GHz during
2012 March and May, respectively.  At those times the array was 
in its C (1.3 cm) and B (5.6 cm) configurations, respectively.  
At 5.307 GHz the phase center was at $\alpha(2000) = 17^h~ 46^m~ 09\rlap.^s0$;
$\delta(2000)$ = $-$28$^\circ~ 42'~ 40.0''$, while at 20.943 GHz, we 
made a small mosaic of three positions covering the 
entire molecular core of G0.253$+$0.016. For both observations the absolute 
amplitude calibrator was 1331+3030 and
the phase calibrator was J1744-3116. 

The digital correlator of the JVLA was configured in 18 spectral windows of 
128 MHz width divided 
in 64 channels of spectral resolution. The total bandwidth for 
both observations was about 2.3 GHz in dual-polarization mode.

The data were analyzed in the standard manner using the
CASA (Common Astronomy Software Applications) and AIPS
(Astronomical Image Processing System) packages of NRAO. 
In both observations, we used the ROBUST parameter of CLEAN set to 0, 
in an optimal compromise between sensitivity and
angular resolution. To construct the continuum in both bands, we only 
used the line-free channels and a bandwidth of 1.5 GHz.
At 5.6 cm, the resulting image rms was 0.24 mJy beam$^{-1}$ at an 
angular resolution of $2\rlap.{''}20 \times 0\rlap.{''}70$ with PA = $-3.4^\circ$.
On the other hand, at 1.3 cm,  the resulting image rms was 
0.045 mJy beam$^{-1}$ at an angular resolution 
of $2\rlap.{''}12 \times 0\rlap.{''}68$ with PA = $166.5^\circ$.

\begin{figure}
\centering
\includegraphics[angle=0,scale=0.6]{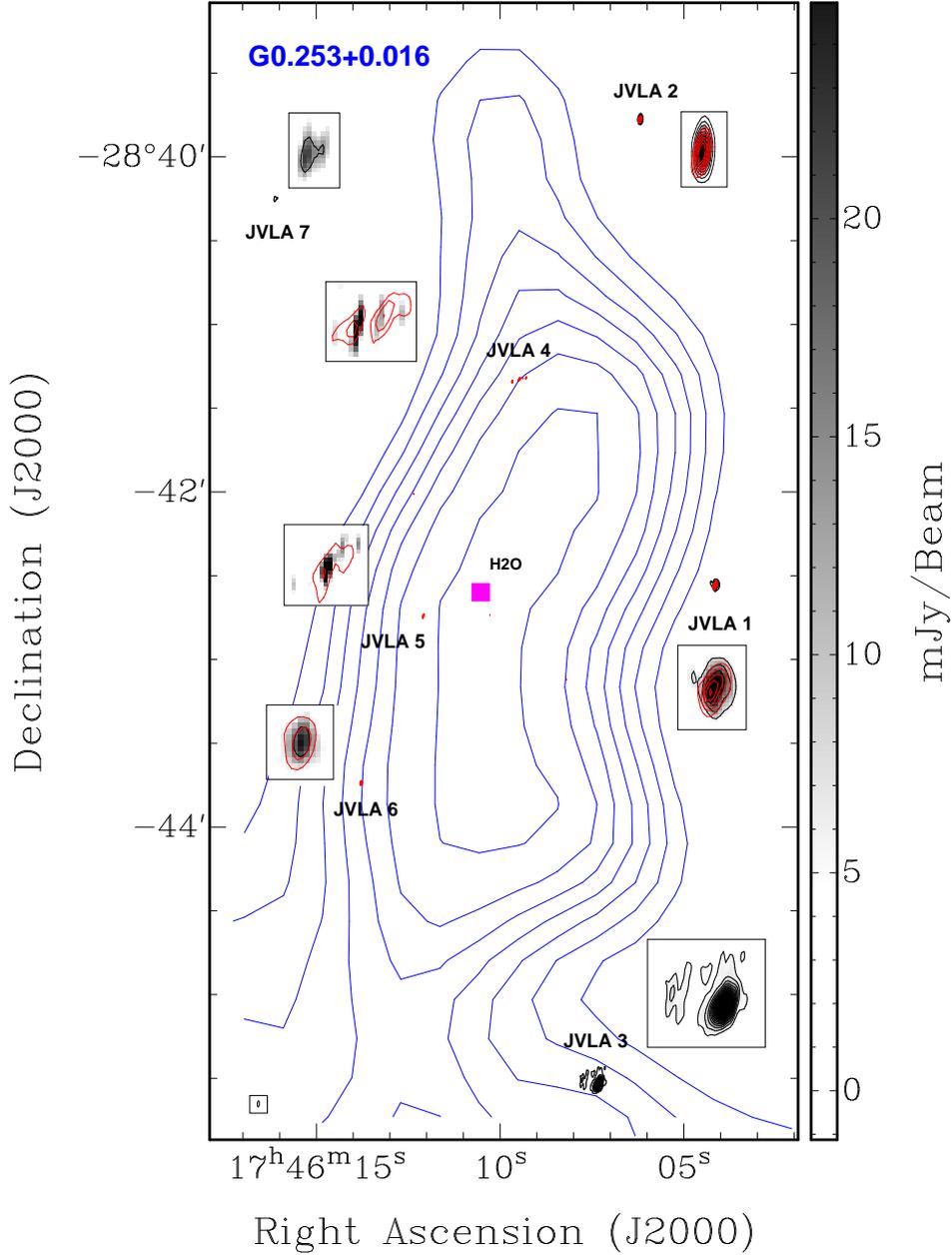}
\vskip-1.0cm
\caption{\small 1.3  (red) and 5.6 (black)  cm continuum 
contour images overlaid on a Herschel SPIRE 250 $\mu$m
image of G0.253$+$0.016 (blue contours). 
The blue contours range from 45\% to 100\% of the peak emission, in steps 
of 9\%. The emission peak for the far-infrared emission is 255 Jy beam$^{-1}$.
The red contours range from 42\% to 90\% of the peak emission, in steps of 5\%. 
The emission peak for the 1.3 cm emission is 1.1 mJy beam$^{-1}$.
The black contours range from 10\% to 90\% of the peak emission, in steps 
of 5\%. The insets show zoom-in images of the radio
sources. The emission peak for the 5.6 cm emission is 25 mJy beam$^{-1}$.
The JVLA contours are not corrected for the response of the primary beam.
The half-power contour of the synthesized beam of the 5.6 cm image is 
shown in the bottom left corner. The grey-scale bar on the right indicates the 
density flux scale of the 5.6 cm image in mJy beam$^{-1}$. The pink square
marks the position of the weak H$_2$O maser in the region.}
\label{fig1}
\end{figure}
\pagebreak

\begin{deluxetable}{l c c c c c c c}
\tablecaption{Parameters of the JVLA sources detected at 1.3 cm}
\tablehead{                        
\colhead{}                        &
\multicolumn{2}{c}{Position} &
\colhead{}                              &
\multicolumn{3}{c}{Deconvolved size$^b$} &        \\
\colhead{}   &
\colhead{$\alpha_{2000}$}          &
\colhead{$\delta_{2000}$}           &
\colhead{Flux Density$^a$ }       &                            
\colhead{Maj.}  &
\colhead{Min.}  &
\colhead{P.A.}  & \\
\colhead{Source}                              &
\colhead{(h m s) }                     &
\colhead{($^\circ$ $^{\prime}$  $^{\prime\prime}$)}              &
\colhead{(mJy)}  & 
\colhead{($^{\prime\prime}$)}  &
\colhead{($^{\prime\prime}$)}  &
\colhead{($^\circ$)} &
}
\startdata

JVLA 1   &   17 46 04.152 & $-$28 42 33.48  &  5.5 $\pm$ 0.3 &  1.6 $\pm$ 0.2 &  1.1 $\pm$ 0.2 & 132 $\pm$ 13\\
JVLA 2   &   17 46 06.194 & $-$28 39 46.65  &  5.8 $\pm$ 0.4 &  1.8 $\pm$ 0.1  &  0.8 $\pm$ 0.1 & 163 $\pm$ 2\\
JVLA 4   &   17 46 09.520 & $-$28 41 19.89  &  4.7 $\pm$ 0.5 &  7.6 $\pm$ 0.3  &  2.1 $\pm$ 0.3 & 104 $\pm$ 6\\
JVLA 5   &   17 46 12.065 & $-$28 42 44.27  &  2.1 $\pm$ 0.7 &  5.2 $\pm$ 0.2  &  2.2 $\pm$ 0.2 & 103 $\pm$ 18\\
JVLA 6   &   17 46 13.789 & $-$28 43 44.22  &  2.1 $\pm$ 0.4 &  1.4 $\pm$ 0.4  &  0.8 $\pm$ 0.4 & 44 $\pm$ 33\\
JVLA 7   &   17 46 16.102 & $-$28 40 14.96  &      $\leq$ 0.4  &            -   &                        - &      -\\
\enddata
\tablecomments{
(a): Total flux density corrected for primary beam response.\\
(b): These values were obtained from the task JMFIT of AIPS.}
\end{deluxetable}

\begin{deluxetable}{l c c c c c }
\tablecaption{Parameters of the JVLA sources detected at 5.6 cm}
\tablehead{                        
\colhead{}   &
\multicolumn{2}{c}{Position} & \\
\colhead{}   &
\colhead{$\alpha_{2000}$}          &
\colhead{$\delta_{2000}$}           &
\colhead{Flux Density$^a$ }       &                            
\colhead{Spectral}  & \\ 
\colhead{Source}                              &
\colhead{(h m s) }                     &
\colhead{($^\circ$ $^{\prime}$  $^{\prime\prime}$)}              &
\colhead{(mJy)}  & 
\colhead{Index$^b$} & 
}
\startdata

JVLA 1   &   17 46 04.140 & $-$28 42 33.32  &  19.9 $\pm$ 1.4  &   $-$0.9$\pm$0.1 \\
JVLA 2   &   17 46 06.186 & $-$28 42 46.52  &  24.4 $\pm$ 0.8   &  $-$1.0$\pm$0.1  \\
JVLA 3   &   17 46 07.315 & $-$28 45 31.99  &  159.2 $\pm$ 3.0 &  $\sim$0$^c$ \\
JVLA 4   &   17 46 09.469 & $-$28 41 19.30  &   2.5 $\pm$  0.2 &    $+$0.5$\pm$0.1 \\
JVLA 5   &   17 46 12.075 & $-$28 42 44.08  &   1.5 $\pm$  0.5 & $+$0.2$\pm$0.3 \\
JVLA 6   &   17 46 13.779 & $-$28 43 44.18  &   3.2 $\pm$  0.8 & $-$0.3$\pm$0.2 \\
JVLA 7   &   17 46 16.102 & $-$28 40 14.96  &   4.7 $\pm$  1.2 & $\leq$$-$1.8$\pm$0.8 \\

\enddata
\tablecomments{
(a): Total flux density corrected for primary beam response.\\
(b): From the 5.6 cm and 1.3 cm data.\\
(c): See text.}
\end{deluxetable}

\section{Interpretation}

In a region of $\sim 3' \times 6'$ centered on G0.253+0.016
we detected seven sources at 5.6 cm with flux densities going from
1.5 to 159.2 mJy (see Fig. 1 and Table 2). Following Anglada et al.
(1998) we estimate that N$\simeq$0.1 background sources are expected 
\sl a priori \rm in this solid angle at 5.6 cm with flux densities
equal or above 1.5 mJy.
We then conclude that all seven sources are most probably related to the
region studied. Three of the sources, JVLA 4, 5, and 6 are located, at
least in projection, inside G0.253+0.016, and have not been reported
previously in the literature. However, given the complexity of the
galactic center region we cannot rule out the possibility that the
apparent association is only a line-of-sight coincidence.

\section{Notes on individual sources}

The spectral index $\alpha$ of a given source ($S_\nu \propto \nu^\alpha$,
where $S_\nu$ is the flux density at frequency $\nu$) was calculated
using

$$\alpha = {{ln(S_1/S_2)} \over {ln(\nu_1/\nu_2)}},$$

\noindent where $S_1$ and $S_2$ are the flux densities at
frequencies $\nu_1$ and $\nu_2$, respectively. The error in the
spectral index $\delta \alpha$ was calculated using standard
error propagation theory (Ku 1966) to give:

$$\delta \alpha \simeq {{1} \over {|ln(\nu_1/\nu_2)|}}  
\biggl[\biggl({{\delta S_1} \over {S_1}} \biggr)^2
+ \biggl({{\delta S_2} \over {S_2}} \biggr)^2 \biggr]^{1/2},$$ 

\noindent where $\delta S_1$ and $\delta S_2$ are the flux density errors at
frequencies $\nu_1$ and $\nu_2$, respectively.

\subsection{JVLA 1, 2, and 7}

These three radio continuum sources 
appear in the periphery but outside G0.253+0.016 (see Figure 1).
Given their non-thermal (negative) spectral index (see Tables 1 and 2), 
it is temptying to classify
them as background extragalactic sources, but as noted before, statistically
we do not expect such sources in the field. Furthermore, sources JVLA 1 and 2
are very bright at 5.6 cm (19.9 and 24.4 mJy, respectively) and we tentatively
propose they are related to the galactic center region. 
Alternatively, we may be fortuitously observing in the direction of a
group of background radio galaxies.
Immer et al. (2012) recently detected sources 1 and 2 with the VLA and tentatively interpreted
them as HII regions, but the non-thermal spectrum determined by us rules out this interpretation
(e.g. Rodr\'\i guez et al. 1993).

\subsection{JVLA 3}

This is the brightest source in the group. It was not covered
by our spatial sampling at 1.3 cm, so we cannot give
its spectral index from our data. However, it was already detected
in the centimeter VLA observations of Downes et al. (1979),
Becker et al. (1994), Lazio \& Cordes (2008), and Immer et al. (2012).
These surveys imply that the
source has an approximately flat spectrum in the centimeter regime,
with a flux density in the range of $\sim$120-180 mJy.
Assuming a flux density of 150 mJy at 5.6 cm resulting 
from optically-thin free-free emission
from gas at an electron temperature of $10^4$ K and that the
source is located at the galactic center, at a distance of 8.0 kpc
(Reid 1993; Reid et al. 2009), we estimate that an ionizing photon rate of
$9 \times 10^{47}$ s$^{-1}$ is required to maintain the source. 
This ionizing rate can be provided by an O9.5 V ZAMS star with
a luminosity of $\sim 4 \times 10^4~L_\odot$ (Panagia 1973;
Thompson 1984). 

Lang et al. (1997) detected the H92$\alpha$ recombination line from this
source at a radial velocity of $v_{LSR}$ = 45.2$\pm$1.1 km s$^{-1}$. 
Walsh et al. (1995)
and Pestalozzi et al. (2005) report associated 6.7-GHz methanol maser emission
at a radial velocity of $v_{LSR}$ = 49.2 km s$^{-1}$.
The source is also a bright far-infrared (IRAS 17430$-$2844) and millimeter source
(Di Francesco et al. 2008; Rosolowsky et al. 2010; Bally et al. 2010).
We estimate the IRAS luminosity of the region to be $\sim 2 \times 10^{5}~L_\odot$,
that can be produced by an O6 ZAMS star. The discrepancy between the
stellar types derived from the photoionization and from the IR luminosity
is well-known in the field of massive star formation (e.g. Carral et al. 1999). 
Our image suggests it has an irregular shell-like morphology with an angular
diameter of $\sim10{''}$. As suggested by Immer et al. (2012), most probably it is an
HII region in the vicinity of G0.253+0.016.

\subsection{JVLA 4, 5, and 6}

These three sources are projected against G0.253+0.016 and their thermal
(flat or positive) spectral indices suggest they are free-free emitters,
optically-thin or partially optically-thick. From their 1.3 cm flux densities
we find that B0.5 ZAMS stars could be providing the photoionization.

JVLA 4 shows an elongated morphology in the east-west direction,
suggesting we may be dealing with a thermal jet (e.g. Anglada 1996; Rodr\'\i guez 1997).
If this were the case, JVLA 4 would be the most radio luminous thermal jet
known (see Fig. 9 of Rodr\'\i guez et al. 2008). Also JVLA 5 shows extended
morphology that deserves further study.

JVLA 6 is associated with the ISOGAL source P J174613.8$-$284344
and proposed to be a bright YSO candidate by Felli et al. (2002).
JVLA 6 is also detected in the Spitzer Space Telescope Galactic Center 
(SSTGC) survey as source SSTGC 618018 (An et al. 2011).
Modeling of the observed IR spectral energy distribution of SSTGC 618018
implies a massive young star with mass of $\sim$9 $M_\odot$ and a
luminosity of $\sim2 \times 10^3~L_\odot$ (An et al. 2011), 
suggesting a B3 ZAMS star. The discrepancy between this estimate
and that of our radio data (that suggests a
B0.5 ZAMS star) requires further investigation.
We detected this radio source
in our analysis of the VLA archive data of project AL302, taken on 1993 June 6,
with a flux density of 1.8$\pm$0.2 mJy at 3.6 cm. This flux density
is consistent with interpolation of our new measurements (Tables 1 and 2),
and suggests that the source is steady in time.

\section{Interpretation and Conclusions}

The seven radio continuum sources detected in the G0.253+0.016 region can
be clearly separated in three groups. JVLA 3 is an optically-thin HII
region probably close, but not directly embedded in G0.253+0.016.
Its photoionization can be provided by a B0.5 V star.
The second group is constituted by the sources JVLA 4, 5, and 6.
All appear projected against G0.253+0.016 and have flat or
positive spectral indices. We suggest they are compact HII regions
powered by early B-type stars. This result implies that,
in contrast to previous conclusions, there is active massive
star formation in G0.253+0.016. Finally, the third group is formed
by the sources JVLA 1, 2, and 7. All appear projected in the
periphery of G0.253+0.016 and have non-thermal indices.
We propose that they could be a class of non-thermal sources in the galactic
center region. The understanding of its nature requires further study.

\acknowledgements
L.F.R. and L.Z acknowledge the financial support of CONACyT, M\'exico
and DGAPA, UNAM.

\end{document}